# Using multi-categorization semantic analysis and personalization for semantic search

**Yinglong Ma**
(North China Electric Power University, Beijing 102206, China
yinglongma@gmail.com)

**Moyi Shi**
(North China Electric Power University, Beijing 102206, China
shimoyi@163.com)

**Abstract:** Semantic search technology has received more attention in the last years. Compared with the keyword based search, semantic search is used to excavate the latent semantics information and help users find the information items that they want indeed. In this paper, we present a novel approach for semantic search which combines Multi-Categorization Semantic Analysis (MCSA) with personalization technology. The MCSA approach can classify documents into multiple categories, which is distinct from the existing approaches of classifying documents into a single category. Then, the search history and personal information for users are significantly considered in analysing and matching the original search result by Term Vector DataBase (TVDB). A series of personalization algorithms are proposed to match users' personal information and search history. At last, the related experiments are made to validate the effectiveness and efficiency of our method. The experimental results show that our method based on MCSA and personalization outperforms some existing methods with the higher search accuracy and the lower extra time cost.

**Categories:** H.3.3, H.3.1, H.2.8

## 1 Introduction

As the amount of information on the Web rapidly increases, information resources are collected by diverse strategies with specific algorithms. However, mainstream search engines are often based on keywords and word frequency statistics, and seldom consider information semantics. The search intentions and needs for different types of users will be not the same even if they use the same query keywords [Jansen, 00]. Users therefore have to further browse each of searched items and determine which items are what they need indeed. It will take more time and efforts for users to browse and select information items they want indeed. In the last years, semantic search technologies have received more attention [Guha, 03; Dong, 10]. Compared with the keyword based search, the goal of semantic search is to explore the latent semantics residing in document information and help users find the information items that they want indeed.

Semantic search has demonstrated its potential successfully in recent years. The research work in semantic search could be roughly classified into three aspects. The first aspect is to analyze the semantic relevancy from the context by calculation models. For example, the hidden Markov tree and other mathematical models were used to calculate the tightness between contexts for analysing the relevancy between retrieval results. These methods contribute to the increased coverage of effective retrieval results [Nguyen, 12; Carpineto, 12; Cao, 09; Lai, 11]. Second, many methods were proposed for optimizing retrieval results. Search results are re-ranked by using some methods such as [Lee, 09; Singh, 10]. Other methods are to analyze the affection dependency between users' queries and related documents [Liu, 10; Paltoglou, 10]. They can greatly improve users' search experience. The third is to construct the personalized semantic retrieval systems based on personal information and historical retrieval records. Some methods summarized users' search habits by the historical retrieve records [Wang, 12] and personal information [Teevan, 05; Carmel, 09; Liu, 10; Pang, 11] By personalization, retrieval accuracy can be significantly improved.

However, some open problems remain to be resolved in the field of semantic search. On one hand, collecting and accumulating the related data for personalization are time consuming and arduous by the existing personalization methods based on categorization. Personalization search can be achieved only when more data is accumulated and trained for obtaining personalized categorization knowledge. It will take a long time for users to accumulate personal data. And also, the knowledge obtained by training may not reflect the latent semantics that users require indeed. In existing approaches, documents are often classified by only determining whether they belong to some category rather than multiple categories. We argue that the same documents can be probably semantically reduced to multiple categories at the same time. For example, the content of a document is related to revealing corrupt officials. It can be simultaneously classified into two categories such as "anticorruption" and "politician". On the other hand, most existing approaches for personalization do not consider the different significances in personal information from different perspectives of users. Different types of users possibly have different search preferences even if they use same search keywords. For example, assume that there are two users to use the same string "prices of apples" for searching what they are interested in. They have different personal information. One is a farmer, and the other is an engineer for developing embedded systems. They generally have different search requirements. The former is to search the prices of apples that s/he can eat, and the latter is to search the prices of electronic products with the brand "Apple". How to fully consider the background knowledge about personal information for semantic search is a challenging task. The method for effectively and efficiently extracting personal information is required to re-organize and re-rank searched results.

The goal of this paper is to address the problems mentioned above. We present a novel semantic search approach which combines Multi-Categorization Semantic Analysis (MCSA) with personalization technology. A uniform framework based on MCSA and personalization is proposed for materializing our semantic search process. The MCSA approach can classify documents into multiple categories, which is distinct from the existing approaches classifying documents into a single category. The search history and personal information are significantly considered in analyzing and matching the original search result by Term Vector DataBase (TVDB). A series of algorithms are proposed to match users' personal information and search history. At last, the related experiments are made to validate the effectiveness and efficiency of our method. The experimental results show that our method outperforms some existing methods with the higher retrieval accuracy and the lower extra cost.

The rest of this paper is organized as follows. Section 2 is the related work. Section 3 is the overview of our framework. In Section 4, we discuss multi-categorization semantic analysis. Section 5 is the semantic personalization technology. In Section 6, the related experiments are made for evaluating our method. Section 7 is the conclusion and the future work.

## 2    Related work

The representative work in analyzing semantic relevancy includes the literature [Cao, 09; Lai, 11; Jeffrey, 06; Eberlein, 11]. They mainly focused on calculating the semantic relevancy from the context through novel calculation models. [Lai, 11; Eberlein, 11] quantified the semantic relationship between keywords and results by using some methods such as fuzzy sets and Hidden Markov Tree. [Cao, 09; Jeffrey, 06] determined the similarity and conformity between search results by a learning context. These methods can improve the semantic relevance of the search results and the recall ratio of a search. However, the time-consumption based on these methods will possibly increase a lot.

Some studies paid attention to improving the performance by optimizing the retrieval results [Lee, 09; Singh, 10; Le, 09; Zhao, 11]. [Singh, 10] attempted to improve the user's search experience by re-organizing the results based on a clustering method. [Lee, 09] re-ranked the original results to make them more reasonable by improving the keyword based sorting algorithms. Other work [Le, 09; Zhao, 11; Zhou, 07] put forward a series of ranking strategies for efficient search. However, these methods did not guarantee the search precision, and failed to clarify why the clustering distinction is reasonable.

Many personalization algorithms were proposed for semantic search in recent years [Imielinski, 09; Wang, 12; Brin, 98; Makris, 07; Celik, 05; Carmel, 09; Liu, 10; Pang, 11; Jeon, 08; Micarelli, 07; Ferragina, 08; Ma, 07; Bouras, 09; Sheng, 09]. [Liu, 04] defined the "category labels set" according to users' search history records. It classified users' queries into three categories using the classification algorithm. This method can improve the accuracy of the search results, but the classification completely depends on the pages of ODP (open directory project) and requires users to use the search engine for a long time to accumulate the historical data. [Li, 07] had noticed that the historical data of users' were update frequently, so an "adaptive" approach was proposed to obtain two kinds of personal information from user's historical records. They replaced the cache by using the algorithm similar to LFU. However, because of the uncertainty of the human activity, we cannot make the assumption that the user's preference should change by several different searches in a short time. [Teevan, 05] claimed to establish personal information index in users' desktop with articles, emails, messages and documents. However, searching information from local computers involves personal privacy issues. The information with special formats is difficult to get.

The work most similar to our approach is the Concise Semantic Analysis (CSA) in [Li, 11]. It found a reasonable way to explain a word or a document in a space of concept that are closely related to their category labels. The category labels are used to build a concept space. Both words and documents will be represented as vectors in the space of concepts. It can effectively reduce the dimension of document vectors, and thus reduce the time consumption by using SVM to classify them. However, when the terms' weights in documents are computed, they failed to involve the positive and the negative documents. These limitations obviously impede more accurate categorization to documents, and will bring about a low retrieval precision.

Our method proposed in this paper is distinct from existing approaches. We introduce a complete framework to materialize multi-categorization semantic analysis and personalization. First, the proposed MCSA approach is to classify documents into multiple categories, which is distinct with existing approaches classifying documents into a single category. Second, both users' search history and personal information are used to analyze and match the original search result by term vector database (TVDB) proposed in this paper. Different types of users will be assigned different preferences by weight scoring. To the best of our knowledge, it is the first work that combines multi-categorization semantic analysis with personalization for semantic search.

## 3  Overview of framework

The framework for the overall process is shown in Fig.1. Specifically speaking, it deals with the following steps.
1) We first extract texts from the pages crawled, and pre-process them to form intermediate documents. Some of these intermediate documents are further processed by the process of MCSA. The others are sent to the indexer and wait to participate in the index creation.
2) The MCSA process starts. Intermediate documents are structured as term vectors, which are stored in the term vector database (TVDB). Term vectors are sent to a SVM classifier for further processing.
3) The indexer generates the index by combining intermediate files with the classification results. Until now, the earlier stage of the work has been accomplished.
4) Users' queries can be parsed and further sent to the TVDB interface. Searching can be made by inquiring TVDB and matching categories. And then, the related information is sent to index searcher and the original set of search results is obtained.
5) The original set of search results is further processed by analyzing users' references according to their personal information and historical records. Here, analyzing personal information also needs to access TVDB. The search results will be further re-ranked according to users' references. At last, the final search results are returned to users.

In the framework, our main contributions are to design and implement the two core components. The first component is MCSA for constructing TVDB and classifying Web pages. The second is the semantic personalization technology (SPT) for optimizing the original search results by matching personal information and historical information.

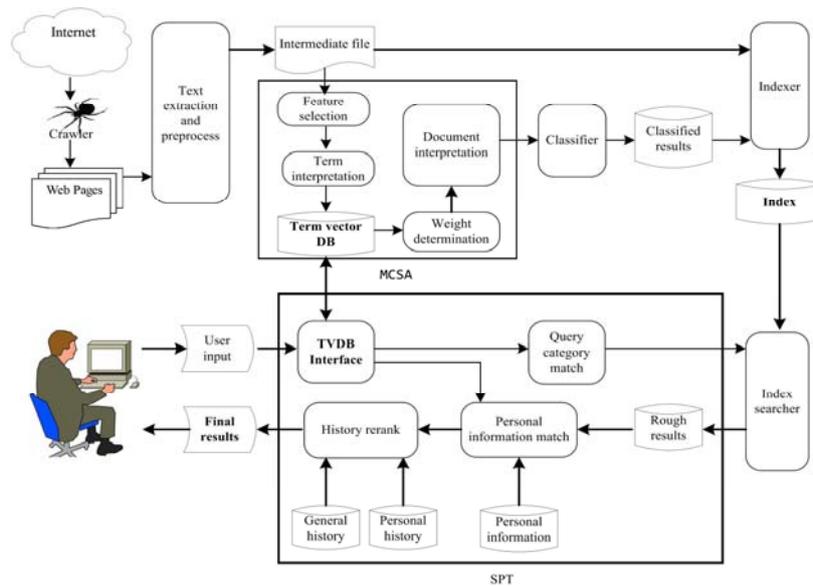

*Figure 1: Overview of framework for multi-categorization semantic analysis and personalization*

## 4 Multi-categorization semantic analysis

This paper uses the TFIDF algorithm to calculate the terms' weights. The SVM technology is used to classify documents. Our approach makes fully use of the data generated in semantic analysis, and has some advantages as follows. 1) The vectors are built based on the space of concepts extracted from category labels. Generally, the dimension is substantially less than the conventional method like bag of words. 2) Due to the low dimension of concept space, we only need to remove the noise words for improving the performance in the stage of feature selection without the need to perform a series of complex operations of reducing the dimensions. 3) Some term vectors with certain semantic relevancy are add into TVDB. Terms possibly are words, phrases or other indexing entities to identify the content of a text document, but only words are used as terms in this paper. The constructed term vectors can effectively support the personalization algorithms at the later stage.

### 4.1 Building concept space

A concept space consists of a concept vector where concepts are derived from the category labels of corpus. A concept space could represent the entire category information of a corpus. It is feasible to interpret words and articles in a concept space. We directly derive concepts from category labels and construct a concept space. In this way, a category label corresponds to a concept, which can be applied to all kinds of corpus.

For example, we use a corpus with 8 categories whose label set is {*"education"*, *"sport"*, *"art"*, *"medical"*, *"transportation"*, *"economy"*, *"science"*, *"military"*}. Each of these labels has clear meaning and will be derived as a concept. We therefore get a concept space with 8 dimensions. Each dimension corresponds to a concept. Thus we can interpret whether a word/document is related to a concept.

### 4.2    Modeling term vectors

After building a concept space, we calculate the document vectors. Considering documents are composed of words, we will calculate the tightness between each term and each concept before generating the document vectors. Then we can model the term vectors. We calculate the tightness $w(c_i, t_j)$ between term $t_j$ and concept $c_i$ using the formula (1), where $d_k$ is the document that contains term $t_j$ and belongs to $c_i$, and the *tf* represented the terms frequency.

$$w(c_i, t_j) = \sum_k H(c_i, d_k) \frac{\log(1 + tf(d_k, t_j))}{\log(1 + length(d_k))}$$

$$H(c_i, d_k) = \begin{cases} 1 & d_k \in c_i \\ 0 & \text{otherwise} \end{cases} \quad (1)$$

We treat the tightness $w(ci, tj)$ as a dimension of a term vectors by calculating the tightness between each term and each concept. That means that each dimension in a term vector represents a relationship between this term and each concept. In this way, we will obtain a term vector in this concept space. Suppose that we have a concept space with three concepts {*"music"*, *"sports"*, *"education"*}, then we use term *"piano"* to calculate its tightness with each of the three concepts. The term *"piano"* has the closest relevancy with *"music"*, and has the loosest relevancy with *"sports"*. We further suppose that their corresponding scores are {*"9"*, *"2"*, *"5"*}. Then, if we acquire a term vector of *"piano"*, its dimensions are {*"9"*, *"2"*, *"5"*}, corresponding to the concepts {*"music"*, *"sports"*, *"education"*}. The advantage of our method is that we can quantify the conceptual meaning.

After modeling term vectors, we make the normalization by using formula (2), which can make each of dimension values of term vectors falls into [0, 1].

$$w'_{ij} = \frac{w(c_i, t_j)}{\sum_i w(c_i, t_j)} \quad (2)$$

Now, we completed the work of structuring term vectors. All the terms extracted from document corpus are stored in TVDB, which will be used for analysis and use in the later stage.

### 4.3    Generating the document vectors

The central values of all the valid term vectors contained in documents are not just the average value of all the term vectors because different words in documents possibly have different significances, which means that a word in different articles possibly have different weights.

In order to measure the significances of each word, this paper uses TFIDF algorithm to calculate the words weights:

$$w_{t_j} = \text{TFIDF}(t_j, d) = tf(t_j, d) \cdot \lg \frac{|D|}{|\{d_k : t_j \in d_k\}|} \quad (3)$$

Where $D$ is the set of documents in entire corpus and $d$ is the current document which contains term $t_j$.

The TFIDF weight algorithm does not involve the positive and negative documents, so it can be used in multi-categorization. For a document $d_k$, dimensions can be calculated by formula (4).

$$wd(c_i, d_k) = \sum_{t_j \in d_k} w(c_i, t_j) \cdot \text{TFIDF}(t_j, d_k) \quad (4)$$

The formula (4) can be expressed as a vector form as follows.

$$\vec{d_k} = \sum_{t_j \in d_k} \text{TFIDF}(t_j, d_k) \cdot \vec{t_j} \quad (5)$$

Similarly, it also has the corresponding normalized formula.

$$w_i' = \frac{w_i}{\sqrt{\sum_{j=1}^{|n|} w_j^2}} \quad (6)$$

Up to now, a document has been interpreted as a vector in the space of concept. The number of dimensions depends on the number of the concepts in the concept space. In this paper, the dimensions of a document vector represent the tightness between the document and concepts in the concept space.

## 5 Semantic personalized technology

### 5.1 Selection and analysis of users' information Future Work

Personal information commonly includes occupation, hobbies, gender, age, character and other characteristics that influence the users' search intentions. However, due to the personal privacy, the way to obtain the information is limited. Only three kinds of personalized information can be used in this paper. They include occupation, hobbies and gender, whose significances are different.

In order to determine the weights of significances, the weights are empirically tuned as follows: *hobbies: 0.5, occupation: 0.3, gender: 0.2*.

### 5.2 Semantic matching

#### 5.2.1 Matching queries

We directly use the term vector database for matching the user's query keywords. The merit is that each of the term vectors has a strong semantic association. Suppose that the user's query is composed of the keyword set *K* after removing the noise words, where *K*={$k_1, k_2,...k_n$}. And the keyword set corresponds to the term vector *T*, where *T*={$T_1, T_2,...T_j$}. If $k_i$ does not exist in the term vector database, the set $T_i$ is a zero vector. If the number of the concepts in the original concept space is *m*, the entire query vector *Q* can be represented as:

$$Q = \sum_{i=1}^{n} T_i = (\alpha_1, \alpha_2, \ldots, \alpha_m) \quad (7)$$

According to [Jansen, 00], a query can be possibly divided into three categories. So we get the vector of top three queries, whose dimensions are the largest. Suppose the vector is ($\alpha_1, \alpha_2, \alpha_3$), and their corresponding categories are $c_1, c_2, c_3$, we can get the three types of documents from the classified target document. And the category weight vector *W* can be generated according to formula (8).

$$W = (\alpha_1, \alpha_2, \alpha_3) / (\alpha_1 + \alpha_2 + \alpha_3) = (w_1, w_2, w_3) \quad (8)$$

We sort the three category documents according to the size of the dimension value *W*, and get the original retrieval result set.

### 5.2.2 Matching personal information

The purpose of personal information matching is to structure two category weight vectors $W_p$ and $W_h$. The personalization algorithm is based on vectors $W_p$ and $W_h$. The method of structuring $W_p$ and $W_h$ is similar with the query vector. The difference is that it does not use the user's query content, but the personal information.

### 5.3 Personalization algorithm

Lucene scoring algorithm is widely acknowledged as a good algorithm for fundamental scoring. We use Lucene scoring algorithm to calculate a score. As a base score, it will be combined with the personalization algorithm for calculating a new score.

A weight vector $W_p$ = ($w_1, w_2, w_3$) corresponds to the category concept {*c1, c2, c3*}. At the same time, a hobby weight vector $W_h$= ($v1, v2, v3$) corresponds to the category concept {*c4, c5, c6*}. Suppose that the concept set *C* = {*c1, c2, c3*} ∪ {*c4, c5, c6*} for each document in the original result set. If the user's information belongs to the concept set *C*, then we use the personalization algorithm to generate a new score for a document by formula (9).

$$newscore = score \cdot (1 + 0.5w_i + 0.3v_i + 0.2s) \cdot (1 + \frac{topscore - score}{topscore - lastscore}) \quad (9)$$

Where *topscore* is the highest score in the original result set and the *lastscore* is the lowest score. The $w_i$ and $v_i$ corresponding to the concept $c_i$ are respectively the dimensions of vectors $W_p$ and $W_h$. If the document is relate to gender, *s*=1. Otherwise, *s*=0.

### 5.4 Historical data optimization algorithm

When a search is done, the clicked keywords and the number of clicks will be also recorded. We call them historical records. If a document has been clicked by a large number of users, we call it hot links. The historical data are dynamically changing and it will affect the user's potential search intention.

In this paper, both the numbers of user's history click records and the hot links clicks are considered in the historical data optimization algorithm. If the document $d$ is in the user's historical click records, or is a hot link, we use formula (10) to generate the new ranking $r'(d)$. Otherwise, we skip this step.

$$r'(d) = \left\lfloor \frac{\sqrt{r(d)}}{s \cdot \log_2(2+n_1) + h \cdot \log_2(2+n_2)} \right\rfloor \quad (10)$$

Where $n_1$ is the number of the historical clicks and $n_2$ is the number of the hot link clicks. If this document is in the user's historical records, we set $s=1$. Otherwise, $s=0$. Similar is to $h$.

## 6 Experiments and performance evaluation

### 6.1 Data sets

In order to validate the retrieval effectiveness, this paper uses the Sogou news corpus1 and some Web page from Yahoo directory as the data sets. After dealing with these data corpora, we use the evaluation methods from [Dou, 09; Tamine-Lechani, 10] to make the related experiments. Meanwhile, we will compare our methods with the Lucene retrieval method and Yahoo Directory Online Category Search, respectively.

### 6.2 Evaluation indicators

In this paper, we use four evaluation indicators to validate our retrieval method presented in this paper.

1) Retrieval accuracy assessment

For each query, we extracted the top 10 result documents as experimental data. Among the 10 documents, the number of the documents satisfying user's requirements is denoted as $D_r$, so we can calculate the retrieval accuracy with the following formula:

$$R = \frac{D_r}{10} \quad (11)$$

2) Integrated retrieval efficiency evaluation

Discounted cumulative gain (DCG) is an indicator that can measure the overall retrieval effect from the search engine algorithms. It is built on the basis of PI. PI is a scoring method that users can assign a score to each document extracted from the top 10 retrieval results to determine whether the document is matching the queries. In general, the assignments 2, 1 and 0 respectively mean that the document is good, fair and bad. With these data, we can get the *DCG* using the following formula.

---

[1] www.sogou.com/labs/dl/c.html

$$DCG_P = rel_1 + \sum_{i=2}^{p} \frac{rel_i}{\log_2 i} \quad (12)$$

Where $i$ is the rankings, $rel_i$ is the PI score given by users and $p$ is the number of the document, here $p$ is set as 10.

3) Sorting effect assessment

$nDCG$ is a sorting effect evaluation index based on $DCG$. It is only assess the pros and cons under the same result.

$$nDCG_p = \frac{DCG_p}{IDCG_p} \quad (13)$$

In the experiment, we selected the top 10 documents for sorting and calculating the $nDCG$ value.

4) Time complexity evaluation

Because the Yahoo Directory online retrieval has the network latency, we only compare the Lucene retrieval method with our method. The indicator of time complexity evaluation measures the time consumed by using the two methods.

### 6.3 Experiments results and analysis

Based on the evaluation index above, we invited seven students as search users to make the related search experiments. In this experimental, *"Lucene"* and *"Yahoo"* respectively represent the original search results of using Lucene and the Yahoo Directory online. *"history optimization"* and *"personalized"* respectively represent the experimental results using the historical data optimization algorithm and the personalized algorithm. *"comprehensive"* represents the results using the combination of the two algorithms above.

#### 6.3.1 Retrieval accuracy

According to the calculate formula (11), we use the same queries with different search methods for evaluating average retrieval accuracy, the experimental results are shown in Table 1. The "average" refers to the average value of results obtained by seven users.

|         | Lucene | Yahoo | History | Personalized | Comprehensive |
|---------|--------|-------|---------|--------------|---------------|
| User1   | 0.312  | 0.667 | 0.441   | 0.680        | 0.689         |
| User2   | 0.464  | 0.650 | 0.510   | 0.639        | 0.701         |
| User3   | 0.315  | 0.579 | 0.347   | 0.568        | 0.623         |
| User4   | 0.440  | 0.644 | 0.522   | 0.651        | 0.678         |
| User5   | 0.556  | 0.700 | 0.556   | 0.734        | 0.734         |
| User6   | 0.383  | 0.516 | 0.401   | 0.621        | 0.668         |
| User7   | 0.453  | 0.671 | 0.492   | 0.749        | 0.755         |
| Average | 0.418  | 0.629 | 0.467   | 0.663        | 0.693         |

*Table 1: Average retrieval accuracy*

Table 1 provides the detailed data returned by each of the search methods under different user's experience. We can find that, compared with Lucene search, the personalized, history optimization and comprehensive search have a significant improvement in retrieval accuracy. Using one of the methods "history" and "personalized" may not obviously improve retrieval accuracy. For example, compared with Yahoo method, the history method has lower values. However, their comprehensive method does have an obvious improvement in retrieval accuracy. The comprehensive method combining with history and personalization has the highest value among these methods. This means that our method of combing MCSA with SPT has outperformed other methods.

Because retrieval accuracy only reflects the average accuracy extracted from the top 10 retrieval results, without considering the merits of the integrated retrieval. In the next, we made experiments about integrated retrieval efficiency.

### 6.3.2 Integrated retrieval efficiency

The experiments related to integrated retrieval efficiency is shown in Table 2. It is not difficult to find that our comprehensive method in integrated retrieval index (DCG values) is superior to the other methods including the Lucene, Yahoo, history and personalized methods. Especially for the keywords with polysemy (like query 2) or involved in different fields (like query 1), our comprehensive method's retrieval performance has outperformed other retrieval methods. This means that our comprehensive method is experimentally effective because it benefits from the efficient multi-categorization semantic analysis and personalization. When dealing with the semantic fuzzy queries, our method also has higher identification accuracy.

|  | Lucene | Yahoo | History | Personalized | Comprehensive |
|---|---|---|---|---|---|
| Apple | 3.533 | 5.953 | 3.867 | 7.321 | 8.243 |
| Doctor | 2.668 | 3.476 | 2.668 | 5.073 | 5.037 |
| Chinese food | 5.571 | 6.380 | 6.876 | 5.601 | 6.906 |
| Race car | 6.279 | 6.438 | 6.685 | 6.754 | 7.185 |
| Beauty | 4.247 | 4.363 | 4.435 | 7.382 | 7.782 |
| House | 4.445 | 4.875 | 5.301 | 6.770 | 7.053 |
| Football | 5.754 | 6.449 | 6.056 | 5.902 | 6.885 |
| Milan | 5.836 | 6.641 | 5.836 | 8.797 | 8.797 |
| Fruit | 3.430 | 5.566 | 4.330 | 8.146 | 8.320 |
| Computer | 6.236 | 6.903 | 7.161 | 7.516 | 7.572 |
| Average | 4.798 | 5.817 | 5.339 | 7.054 | 7.267 |

*Table 2: Three retrieval methods of its DCG value in ten different queries*

### 6.3.3 Sorting effect (nDCG indicator)

We calculate the nDCG values with the same queries set (containing 10 keywords). The experimental results are shown in Table 3. In table 3, we can find, compared with Lucene and Yahoo, no matter which method including history, personalized or comprehensive is used, it always shows a stable sorting effect.

Furthermore, the comprehensive method outperforms other methods including Lucene, Yahoo, history and personalization method, it shows a great performance.

For the keywords with polysemy (like query 2) or involved in different fields (like query 1 and query 5), our method's retrieval performance has the highest performance. When the keywords have clear meanings (i.e., user's intention is very clear), or have close relevancy with certain categories, the effect of users' personal information matching is not significant.

|              | **Lucene** | **Yahoo** | **History** | **Personalized** | **Comprehensive** |
|--------------|-----------|-----------|-------------|------------------|-------------------|
| Apple        | 0.36      | 0.82      | 0.44        | 0.81             | 0.89              |
| Doctor       | 0.39      | 0.55      | 0.39        | 0.72             | 0.72              |
| Chinese food | 0.63      | 0.87      | 0.77        | 0.65             | 0.81              |
| Race car     | 0.84      | 0.81      | 0.88        | 0.84             | 0.97              |
| Beauty       | 0.72      | 0.78      | 0.84        | 0.93             | 0.96              |
| House        | 0.76      | 0.82      | 0.78        | 0.87             | 0.93              |
| Football     | 0.92      | 0.89      | 0.94        | 0.92             | 0.94              |
| Milan        | 0.81      | 0.91      | 0.81        | 0.89             | 0.89              |
| Fruit        | 0.51      | 0.80      | 0.76        | 0.84             | 0.84              |
| Computer     | 0.89      | 0.89      | 0.91        | 0.86             | 0.95              |
| Average      | 0.682     | 0.818     | 0.754       | 0.823            | 0.890             |

*Table 3: Three retrieval methods of its nDCG value in ten different queries*

### 6.3.4 Time complexity

Time complexity is computed according to the waiting time from inputting a query to returning the results. Yahoo Directory online was not compared with our approach because it has a significant network delay when we have to access its server. The experiments were made on a PC with Frequency 2.0Hz and Memory 2GB. The experimental results about average retrieval time are shown in Table 4. Due to the additional optimization work, our method has an increased extra time consumption compared with Lucene according to table 4. Both the two methods consumed a little time: one is 0.343 second, and the other is 0.669. The extra time cost is 0.326s. We argue that such extra time consumption is tolerant for performing more effective semantic search, compared with other methods.

|                        | **Lucene** | **Our method** |
|------------------------|-----------|----------------|
| Average retrieval time | 0.343s    | 0.669s         |

*Table 4: Time complexity comparison*

In brief, we use four indicators to evaluate our method. The related experimental results show that our approach is effective and efficient by combing MCSA and personalization, and outperforms other methods with the lower extra time cost.

# 7    Conclusions

In this paper, we proposed an approach for semantic retrieval based on the multi-categorization semantic analysis and the personalization technology. First, the document was classified using the improving categorization semantic analysis technology, and then established the term vectors database (TVDB). Second, the user's personal information and historical retrieve records were considered in analyzing and matching the original retrieval result set by TVDB in order to optimize the search results. The experimental results show that our method can significantly improve the semantic relevancy and the retrieval accuracy with the lower extra time cost.

In the future work, we focus on using TVDB to calculate the semantic similarity of the keywords. By doing this, we could get some relevant documents which have the synonymic semantics. And this could further improve the accuracy of the semantic searching.

**Acknowledgments**

This work is partially supported by National Natural Science Foundation of China under (61001197, 61372182), National Key Basic Research Program of China (2009CB320704), and the Fundamental Research Funds for the Central Universities.